\documentclass[11pt,a4paper]{article}
\usepackage{times}
\usepackage{dsfont}
\usepackage{fullpage}
\usepackage{titlesec}
\usepackage[bookmarks=false]{hyperref}
\usepackage[normalem]{ulem}
\usepackage{graphicx,cite}
\usepackage{amssymb,amsmath,amsthm}
\usepackage{amssymb,bm}
\usepackage{array}
\usepackage{amsfonts}
\usepackage{enumitem}
\usepackage{color}

\usepackage{ifthen}

\usepackage{setspace}
\usepackage{amsmath,amsfonts}
\usepackage{amssymb}
\usepackage{graphicx}

\usepackage{amsmath,amscd,amssymb,amsgen,amsfonts,amsbsy,amsthm}
\usepackage{mathrsfs}
\usepackage{bm}
\usepackage{bbm}
\usepackage{color}
\usepackage{textcomp}
\usepackage{float}
\usepackage{latexsym,graphicx}
\usepackage{url}
\usepackage{todonotes}
\usepackage{authblk}

\theoremstyle{plain}

\newtheorem{theorem}{Theorem}

\newtheorem{proposition}[theorem]{Proposition}

\theoremstyle{remark}

\begin{document}
	\title{A Survey of Stability Results for Redundancy Systems}
	\author[1,3]{E. Anton}
	\author[1,2,3,4]{U. Ayesta}
	\author[5]{M. Jonckheere}
	\author[1,3]{I.M. Verloop}
	\affil[1]{CNRS, IRIT, 2 rue Charles Camichel, 31071 Toulouse, France}
	\affil[2]{IKERBASQUE - Basque Foundation for Science, 48011 Bilbao, Spain}
	\affil[3]{Universit\'e de Toulouse, INP, 31071 Toulouse, France}
	\affil[4]{UPV/EHU, University of the Basque Country, 20018 Donostia, Spain}
	\affil[5]{Instituto de C\'alculo - Conicet, Facultad de Ciencias Exactas y Naturales, Universidad de Buenos Aires (1428) Pabell\'on II, Ciudad Universitaria Buenos Aires, Argentina. }
	\date{}
	\maketitle
%%%%%%%%%%%%%%%%
%%%%%%%%%%%%%%%%

\begin{abstract}
	\normalsize
Redundancy mechanisms consist in sending several copies of a same job to a subset of servers. It constitutes one of the most promising ways to exploit diversity in multi-servers applications. However, its pros and cons are still not sufficiently understood in the context of realistic models with generic statistical properties of service-times distributions and correlation structures of copies. We aim at giving a survey of recent results concerning the stability - arguably the first benchmark of performance - of systems with cancel-on-completion redundancy. We also point out open questions and conjectures.
\bigskip 

\textbf{Keywords:} {redundancy, load balancing, stability}

%\textbf{AMS(2020) subject classification:} Primary 60K25, Secondary 68M20 
\end{abstract}

%Whereas for homogeneous servers the stability region could be reduced when adding redundancy, when adding heterogeneity, the opposite effect can happen. 
%

% Sample
%\KEYWORDS{deterministic inventory theory; infinite linear programming duality;
%  existence of optimal policies; semi-Markov decision process; cyclic schedule}

% Fill in data. If unknown, outcomment the field
%\KEYWORDS{Redundancy model, load balancing, stochastic stability}
%\maketitle
%%%%%%%%%%%%%%%%%%%%%%%%%%%%%%%%%%%%%%%%%%%%%%%%%%%%%%%%%%%%%%%%%%%%%%

% Samples of sectioning (and labeling) in OPRE
% NOTE: (1) \section and \subsection do NOT end with a period
%       (2) \subsubsection and lower need end punctuation
%       (3) capitalization is as shown (title style)./
%
%\section{Introduction.}\label{intro} %%1.
%\subsection{Duality and the Classical EOQ Problem.}\label{class-EOQ} %% 1.1.
%\subsection{Outline.}\label{outline1} %% 1.2.
%\subsubsection{Cyclic Schedules for the General Deterministic SMDP.}
%  \label{cyclic-schedules} %% 1.2.1
%\section{Problem Description.}\label{problemdescription} %% 2.

% Text of your paper here

%%%%%%%%%%%%%%%%%%%%%%%%%%%%%%%%%%%%%%%%%%%%%%%%%%%%%%%%%%%%%%%%%%%%%%%%%%%%%
%%%%%%%%%%%%%%%%%%%%%%%%%%%%%%%%%%%%%%%%%%%%%%%%%%%%%%%%%%%%%%%%%%%%%%%%%%%%%

\section{Introduction}\label{ua:intro}

While there are several variants of redundancy-based systems, the general notion of redundancy is to dispatch multiple copies of each job to a subset of servers and to consider the result of whichever copy completes service first.
By allowing for redundant  copies, the aim is to minimize the system latency
by exploiting the variability in the queue lengths of the different queues. 
The potential of redundancy mechanisms lies in finding
the right trade-off between exploiting variability and  the waste of resources induced by having  redundant copies.

Several empirical (\cite{Ananthanarayanan2012,Ananthanarayanan13,Dean2012,Dean13,Sieber2015,Vulimiri13}) and numerical studies (\cite{Gardner2017,Gardner17b,Joshi17,Shah16,Lee17a}) suggest that redundancy might potentially improve the performance of real-world computer system applications. In particular, 
%for instance
Vulimiri et al. \cite{Vulimiri13} consider a 10 DNS servers system and compare the system where each arriving query dispatches 10 copies to all the 10 DNS servers, to an alternative system where queries are assigned to a single server chosen uniformly at random. The authors observe that the fraction of queries with a service time exceeding 500 ms is reduced by a factor 6.5, and the fraction exceeding 1.5 sec is reduced by a factor 50. Another interesting study is provided by Dean and Barroso \cite{Dean13} who underline that several redundancy techniques are applied in Google's BigTable in order to improve the latency of incoming queries. They show that a redundancy system with two copies %where upon arrival of a query two copies are dispatched into two servers, 
reduces the median response time by 16\% and the 99.9th-percentile of the tail of the response time distribution by nearly 40\% compared to the non-redundant system.

%The general notion of redundancy is that upon arrival each job is dispatched into multiple servers in
%order to exploit the variability of the queue lengths and server capacities in the system. 
Broadly speaking, depending on when replicas are deleted, we can consider two classes of redundancy systems: cancel-on-start $(c.o.s.)$ and cancel-on-completion $(c.o.c.)$. In redundancy systems with $c.o.c.$, once one of the copies has completed service, the other copies are deleted and the job is said to have received service. In redundancy systems with $c.o.s.$, copies are deleted as soon as one copy starts being served, and as a consequence,  $c.o.s.$ does  not waste any computation resources. 

In this survey, we will provide an overview on stability results in redundancy systems. From the point of view of stability, $c.o.s.$ does not have any negative impact, and for this reason  we focus on stability results when $c.o.c.$ is implemented. 
%on the impact of redundancy $c.o.c.$ on the stability region of the system, i.e.,   the set of parameters leading to finite stationary queue-lengths.

%Within this paper, we focus on the stability condition of redundancy models where copies of the job are removed when a first copy completes service, that is,  \emph{cancel-on-complete} ($c.o.c.$) redundancy models. 
Let us illustrate through a simple example how redundancy affects the stability region. Consider a system with $K$ homogeneous servers in which copies of each arriving job are dispatched to $d \leq K$ servers chosen uniformly at random. We assume that jobs arrive according to a Poisson process of rate $\lambda$ and jobs have general service times with unit mean. Without redundancy, i.e. $d=1$, the stability condition under any work-conserving policy is given by $\lambda<\mu K$, where $\mu$ is the capacity of the servers.
Now, let us assume that the service times of copies are i.i.d. and that $d=K$. In this case, the system behaves as a single server system with arrival rate $\lambda$ and server capacity $\mu K$, and the stability condition is again $\lambda<\mu K$.  However, if all the copies had the same service time as the original job (identical copies), servers are synchronized and the instantaneous departure rate is just $\mu$. Therefore, the system behaves as a single server system with arrival rate $\lambda$ and server capacity $\mu$, for which the stability condition is $\lambda<\mu$. This simple example illustrates how the modeling assumptions and the degree of redundancy  can dramatically impact the stability condition of the system.

%Opposite to $c.o.c.$ redundancy models, under \emph{cancel-on-start} ($c.o.s.$) redundancy models, copies of a job are removed as soon as one copy starts being served. Therefore,  $c.o.s.$ exploits the variability of the queue lengths, while there is no waste any computation resources. In spite of this, most of the literature has focused on systems with $c.o.c.$, which exploits both the variability of the queue lengths and the server capacities. Furthermore, this model seems to be more interesting from the application point of view, since the dispatcher is unaware of the state of the servers.

%Within this survey, we focus on the impact of redundancy on the stability region of the system, i.e.,  the set of parameters leading to finite stationary queue-lengths. 

%The literature on stability analysis of redundancy is recent and growing. %We assume the $c.o.c.$ redundancy models and we provide an overview of the existing results on the stability condition.
%In this survey, 
%In this paper, 
%we provide an overview of the existing results on stability conditions for $c.o.c.$ redundancy. 
One of the main lessons we draw from the results available in the literature, is that the stability region depends strongly on the scheduling policy employed at the servers and the correlation structure of copies. Somewhat surprisingly,  we also 
identify situations for which it was shown that adding redundant copies does not reduce the stability region.
Overall, we believe more research is  needed in order to design efficient redundancy algorithms.

The rest of the survey is organized as follows.
%As already mentioned previously, in this chapter we provide an overview of the existing results on the stability condition for $c.o.c.$ redundancy \rv{models}. 
%Most notably, we will identify  set of situations for which it has been shown that redundancy does not reduce stability region. 
%One of the main lessons we draw is that the stability properties of the system depend on the scheduling policy employed at servers, and that more research is needed in order design efficient redundancy algorithms. 
%The rest of the chapter is organized as follows. 
Section~\ref{ua:model_description} describes the main model assumptions and notation, Section~\ref{ua:IID} deals with the case in which the service times of the copies are $i.i.d.$, and Section~\ref{ua:Cor} with identical and correlated copies. In Section~\ref{ua:realted}, we present a brief account of results on redundancy that, even though not directly related to stability, are relevant from the performance point of view. We conclude with Section~\ref{ua:Conclusions} where we discuss several open problems and state various conjectures.

%%%%%%%%%%%%%%%%%%%%%%%%%%%%%%%%%%%%%%%%%%%%%%%%%%%%%%%%%%%%%%%%%%%%%%%%%%%%%%%%
\section{Model Description and Preliminaries}\label{ua:model_description}

We consider a $K$ parallel heterogeneous server system. That is, we have a set of servers $S=\{1,\ldots,K\}$ and server~$s$ has capacity $\mu_s$, for $s\in S$. Jobs arrive to the system according to a Poisson process of rate $\lambda$. Arriving jobs have service times  that are independent across jobs and are identically distributed with mean~1.  
%according to some probability distribution function $f$ with cumulative distribution function $F$.
%\matt{CAREFUL with the confusion between the distribution of copies and the distribution of arriving jobs??, later $X_i$ not iid}

Jobs are  labeled by types $c=\{s_1,\ldots,s_i\}\subset S$, where $i$ is the number of copies and $c$ is the set of servers to which this job will dispatch copies. 
%For a given incoming job, its type $c\subseteq S$ characterizes the set of compatible servers of that job. 
%The job dispatches copies to all its compatible servers.   
We let $\mathcal C$ be the set of all possible types. A job is of type~$c$ with probability $p_c$, where $\sum_{c\in\mathcal C} p_c=1$. 

We consider redundancy models that are $c.o.c.$, that is, as soon as a copy is fully served, the additional copies of that job are removed from the system. This cancellation process induces a correlation in the departure process at the servers. Thus, within a server~$s$ there is a departure of a copy due to the following two events: $i)$ a local copy departs due to completion in server~$s$, or $ii)$ a copy in another server completes that induces a departure in server~$s$. 

\begin{figure*}
	%	\begin{minipage}{0.33\textwidth}
	\centering
	\includegraphics[scale=0.9]{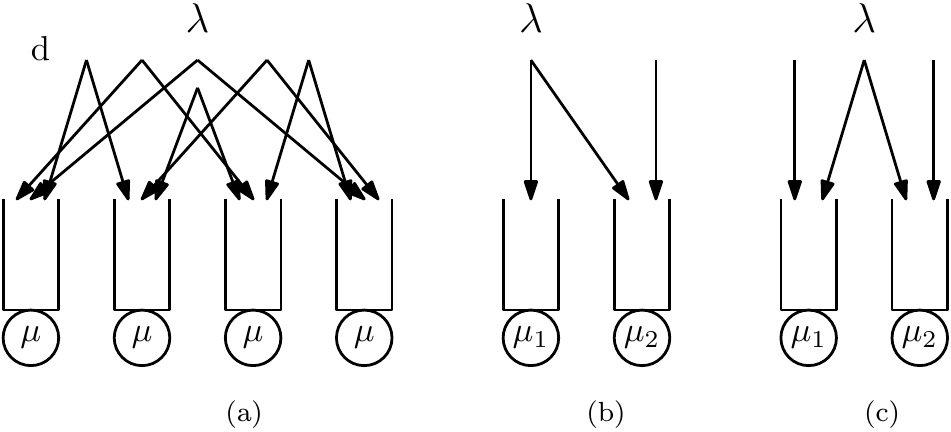}
	%	\end{minipage}
	%	\begin{minipage}{.33\textwidth}
	%		\centering
	%		\includegraphics[scale=0.5]{N_model_fig.pdf}
	%	\end{minipage}
	%	\begin{minipage}{.33 \textwidth}
	%		\centering
	%		\includegraphics[scale=0.5]{Nested_model_fig.pdf}
	%	\end{minipage}
	\caption{(a) The redundancy-$d$ model for $K=4$ and $d=2$. (b) The $N$-model. (c) The $W$-model.}
	\label{ua:fig:red_examples}
\end{figure*}

\textbf{Model Topology.}
A well-known symmetric topology is the one in which each job sends a copy to $d$ out of $K$ servers. In case the server are chosen  uniformly at random, that is, $p_c=1/\binom{K}{d}$, and servers have the same capacity~$\mu$, we refer to this model as the redundancy-$d$ model, see Figure~\ref{ua:fig:red_examples} (a). The number of copies,  $d$,  is referred to as the redundancy degree.

%\textit{Nested model:}  A system is said to have a nested redundancy structure if $\mathcal C$ satisfies the following: for all job types $c,c'\in\mathcal C$, either \emph{i)} $c\subset c'$ or \emph{ii)} $c'\subset c$ or \emph{iii)} $c\cap c'=\emptyset$. First of all, note  that the redundancy-$d$ model does not fit in the nested structure.   The smallest nested system is the so-called $N$-model (Figure~\ref{fig:red_examples} (b)):  this is a $K=2$ server system with types $\mathcal C=\{\{2\},\{1,2\}\}$. Another nested system is the $W$-model (Figure~\ref{fig:red_examples} (c)), that is, $K=2$ servers and types $\mathcal C = \{\{1\}, \{2\}, \{1,2\}\}$. %Under these model jobs of type $\{1,2\}$ always share a server with another job stream. 
%In  Figure~\ref{fig:red_examples} (d), a nested model with $K=4$ servers and $7$ different jobs types, $\mathcal C=\{\{1\},\{2\},\{3\},\{4\},$ $\{1,2\},\{3,4\},\{1,2,3,4\}\}$ is given. This model is referred to as  the $WW$-model. 

Two other examples of redundancy topologies are the so-called  $N$-model and $W$-model, see  Figure~\ref{ua:fig:red_examples} (b) and (c). Both models are non-symmetric, with two servers. The set of possible job types is $\mathcal C=\{\{2\},\{1,2\} \}$ in the $N$-model, and $\mathcal C=\{\{1\},\{2\},\{1,2\} \}$ in the $W$-model.

When no specific structure is assumed, we refer to it in the sequel as a general topology.

%In the remainder of this section, we introduce the different mathematical modeling assumptions made in the literature to analyze redundancy models. 

\
%\matt{MAYBE PUT THE FOLLOWING IN THE INTRO? (here we only focus on the coc case no?)}
%\textbf{Cancellation of copies of a job} might occur either as soon as the first copy of this job enters service, known as cancel-on-start ($c.o.s.$) or as soon as a copy is fully served, known as cancel-on-complete ($c.o.c.$). %Both $c.o.s.$ and $c.o.c$ are interesting since they exploit different aspects of the system: Under $c.o.s$, dispatching copies of a job to multiple servers increases the chance that a job finds a \emph{short} queue and enters service early. Additionally under $c.o.c$, the chance that a job is served in a \emph{fast} server increases.

%Redundancy models are challenging to analyze due to the correlation among the departure processes of the servers induced by the cancellation of the copies. Within a server there is a departure of a copy due to the following two events: $i)$ a local copy departs due to completion, or $ii)$ another copy of a job in the current servers completes service. 

%\rv{rewrite} Dispatching replicas of the same job to several servers increases the chance for one of the replicas to find a short queue and thus start service fast. The ‘cancel-on-completion’ version additionally increases the chance for one of the replicas to have a short execution time (assuming independence among run times on different servers).

\textbf{Scheduling Policy. } A scheduling policy determines how copies are served within each server. As we will see, the choice of the scheduling policy can have a dramatic impact on the stability region.    First-Come-First-Served (FCFS) and Processor Sharing (PS) are widely implemented in real-world computer systems (\cite{HB13}), and are thus 
common policies considered in the literature on redundancy.  Random-Order-of-Service (ROS) is not a common discipline in systems, but as we will see in the ensuing, it yields  very good performance in terms of stability for a redundancy system. These three policies represent the main focus of our survey. 
To the best of our knowledge, other policies such as Last-Come-First-Served (LCFS), Shortest-Remaining-Processing-Time (SRPT), and Least-Attained-Service (LAS) have not been considered so far.

\textbf{Correlation Structure Among Copies.} This describes how the service times of the copies of a given  job are related. 
%, and it  plays an important role when characterizing the instantaneous departure rate of a job in $c.o.c.$ systems.
Formally, the service times $X_1,\dots,X_k$ of the copies of one job can be sampled from a joint   distribution $F(x_1,\ldots,x_k)$. 
% where all $X_k$ are distributed according to the same random variable $X$.
Two extreme cases are   \emph{i.i.d. copies} and    \emph{identical copies}. Under i.i.d. copies,  all copies have independent service times sampled from the same distribution, whereas with  identical copies, all the copies of a job have the same service time. Another interesting framework is the so-called $S\&X$ model introduced in \cite{Gardner17b}. 
Here, the service time of each copy is decomposed into two components; the inherent job size, which is identical for all the copies of a job, and the experienced slowdown on the server it is being served.

\textbf{Existing Stability Results.} Table~1 summarizes the main stability results for $c.o.c.$ redundancy models available in the literature  and discussed in this survey.  The table is organized by scheduling policy, service time distribution, redundancy  topology   and correlation structure. In brackets we specify the additional assumptions that the authors consider in their respective paper.  The term ``red-$d$'' refers to the redundancy-$d$ system and the term ``gen.'' refers to  a general redundancy topology.  
%The general correlation structure also includes i.i.d. copies and identical copies. 
%We have colored in gray the assumptions that are also covered by some more general reference.    

\begin{figure}[bth]
	\centering
	\includegraphics[scale=0.5]{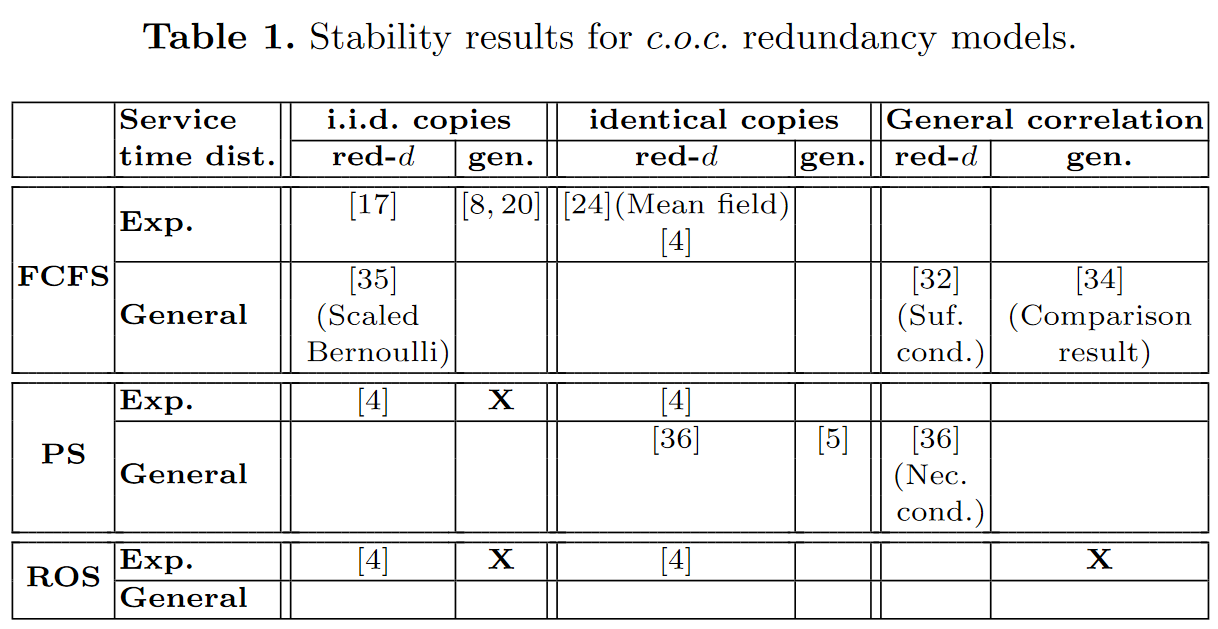}
\end{figure} 

The stability condition when jobs have i.i.d. copies is the main topic of Section~\ref{ua:IID}, first for exponential service times (Section~\ref{ua:IID_exp}) and then  for scaled Bernoulli distributions (Section~\ref{ua:IID_gen}). These are the results in the first two columns of Table~1. Correlated copies are discussed in Section~\ref{ua:Cor}, first for  identical copies (Section~\ref{ua:ident_cop},  middle two columns in Table~1) and then for general correlation structures (Section~\ref{ua:Stab:gen}, last two  columns  of Table~1).  
In Section~\ref{ua:Conclusions}, we discuss open problems and state various conjectures regarding stability conditions. In  Table~1, these conjectures are indicated with a \textbf{X}.

%Finally, in Section~\ref{}, we introduce redundancy literature that deals with other performance measures rather than the stability condition. 

%Furthermore, for both i.i.d. copies and identical copies, the stability region might reduce, but cannot increase. The latter can be explained as follows: when jobs have exponentially distributed service times, the total departure rate is at most $K\mu$, where the total arrival rate is $\lambda$. Hence, a necessary stability condition is $\lambda<K\mu$. 
\section{Independent and Identically Distributed Copies}
\label{ua:IID}

In this section we assume that jobs have i.i.d. copies. 

%\rv{This is repetitive with the section below} We note that under the exponential service times and i.i.d. copies assumptions, the instantaneous departure rate of a job in service is characterized by the sum of the departure rates of all its copies in service.

\subsection{Exponential Service tTmes}
\label{ua:IID_exp}
We first discuss results on   FCFS and    exponentially distributed service times, a setting studied by Gardner et al. \cite{Gardner16,Gardner17} and Bonald and Comte \cite{Bonald17a}. 
It was shown in
\cite{Bonald17a} that  this model fits the framework of Order Independent queues (see \cite[Chapter 2]{Krzesinski2011}), which is a large class of systems that have a product-form steady-state distribution.
This can be seen as follows. 
Since   copies are i.i.d., 
%makes that the instantaneous departure rate of a job in service is characterized by the sum of the departure rates of all its copies in service. 
we can describe the system through the Markovian state descriptor $(c_n, c_{n-1}, \ldots, c_2,c_1)$. Here,   $n$ is the number of jobs in the system, $c_1$ is the type of the eldest job in the system and $c_i$ is the type of the $i$th eldest job. Because of FCFS, the eldest job is served in all of its compatible servers  $c_1$. The $i$-th eldest job is in service at servers $s\in c_i \backslash \cup_{j=1}^{i-1} c_j$, for $i=1,\ldots,n$.
Due to the exponentially distributed service times and i.i.d. copies, the instantaneous departure rate of the i$th$ job is given by the sum of the rates in the servers where the job is in service, that is, $\sum_{s\in c_i\backslash c_1,\ldots,c_{i-1}} \mu_s$. Hence, the total instantaneous departure rate out of state $(c_n, c_{n-1}, \ldots, c_2,c_1)$ is $\sum_{s\in\cup_{j=1}^n c_j} \mu_s$, which depends on the set of classes present in the system, but not on their ordering in the state descriptor, i.e., the so-called order independent property.

The characterization of the steady-state distribution 
facilitates the derivation of performance measures such as the stability condition and mean response times. The proposition below states the stability result for this model.
%The following proposition states the main result on stability: 
%and shows in particular that  In \cite{Gardner16} %Serfozo09  the authors solve the partial balance equations, which is a sufficient and necessary condition for quasi-reversibility (\cite{Kelly1979}) and obtain the steady-state distribution, as well as the stability condition: 

\begin{proposition}[\cite{Bonald17a,Gardner16}]\label{ua:fcfs_iid}
	For a redundancy system with general topology under FCFS with   exponentially distributed service times and i.i.d. copies, the system is stable if for all $C\subseteq\mathcal C$,
	\begin{equation}\label{ua:eq:teo1}
	\lambda \sum_{c\in C} p_c <  \sum_{s\in S(C)} \mu_s,
	\end{equation}
	where $S(C)= \bigcup_{c\in C} \{s\in c\}$. The system is unstable if there exists $\tilde C\subseteq \mathcal C$ such that 
	\begin{equation*}
	\lambda \sum_{c\in \tilde C} p_c >  \sum_{s\in S(\tilde C)} \mu_s.
	\end{equation*}
\end{proposition}

Informally, Equation~(\ref{ua:eq:teo1}) states that the arrival rate to any subset of job types must be less than the total capacity of the associated compatible servers. For exponential service times, this is the \emph{maximum stability} condition, i.e., the system cannot be stable if one of these inequalities were not satisfied. Thus, we conclude that the stability region is not reduced due to adding redundant copies.  The latter might seem counter-intuitive at first, since even if servers waste resources serving copies that are not fully served, the stability condition is as large as if there was no redundancy (see also the simple example in the introduction).

Extending Proposition~\ref{ua:fcfs_iid} to other scheduling policies is an important open problem (see Section~\ref{ua:Conclusions} for more details). To the best of our knowledge, this has only been achieved for the redundancy-$d$ model. In this case, it is easy to see that Equation~(\ref{ua:eq:teo1}) reduces to $\lambda<\mu K$, and it has been shown that this stability condition remains valid when either PS or ROS is implemented.

%Under the i.i.d. copies and exponential service times assumption, the latter result suggest that for any work-conserving  non-preferential scheduling policy the stability condition (\ref{eq:teo1}) is not reduced due to adding redundant copies.
%that is given by Theorem~\ref{ua:fcfs_iid}.

%For the homogeneous servers redundancy-$d$ model, the stability condition is give by $\lambda<\mu K$ where either PS or ROS is implemented.  .

\begin{proposition}[\cite{Anton2019}]\label{ua:ps_ros_iid}
	For the redundancy-$d$ model under either PS or ROS with exponentially distributed service times  and i.i.d.\ copies, the system is stable  when  $\lambda<K\mu$ and unstable when $\lambda>K\mu$.
\end{proposition}

%The authors believe that the above result holds for any non-preferential scheduling policy that treats all job types equally, but did not succeed in obtaining a unifying proof.

Hence, under PS, ROS and FCFS, the redundancy-$d$ model is maximum stable. This however does not hold true in general. 
In the example below (originally in  \cite{Anton2019}),  we describe  priority policy that is  not maximum stable, i.e., the system can become  unstable even though $\lambda<K\mu$. 

%a counterexample to show that if the scheduling policy gives preferences over types, the system under exponential service times and i.i.d. copies can not be maximum stable. 

\ 

\label{ua:ex:counter}
\textbf{Example: Priority Policy.} Consider the redundancy-$d$ system with $K=3$, $d=2$ and $\mu=1$.  There are three different types of jobs: $\mathcal C = \{\{1,2\},\{1,3\},\{2,3\} \}$. In server~1, FCFS is implemented. In server~2 and server~3, jobs of types $\{1,2\}$ and $\{1,3\}$ have preemptive priority over jobs of type $\{2,3\}$, respectively. Additionally, within a type, jobs are served in order of arrival.

\begin{figure}[hbt]
	\begin{center}
		\includegraphics[scale=0.5]{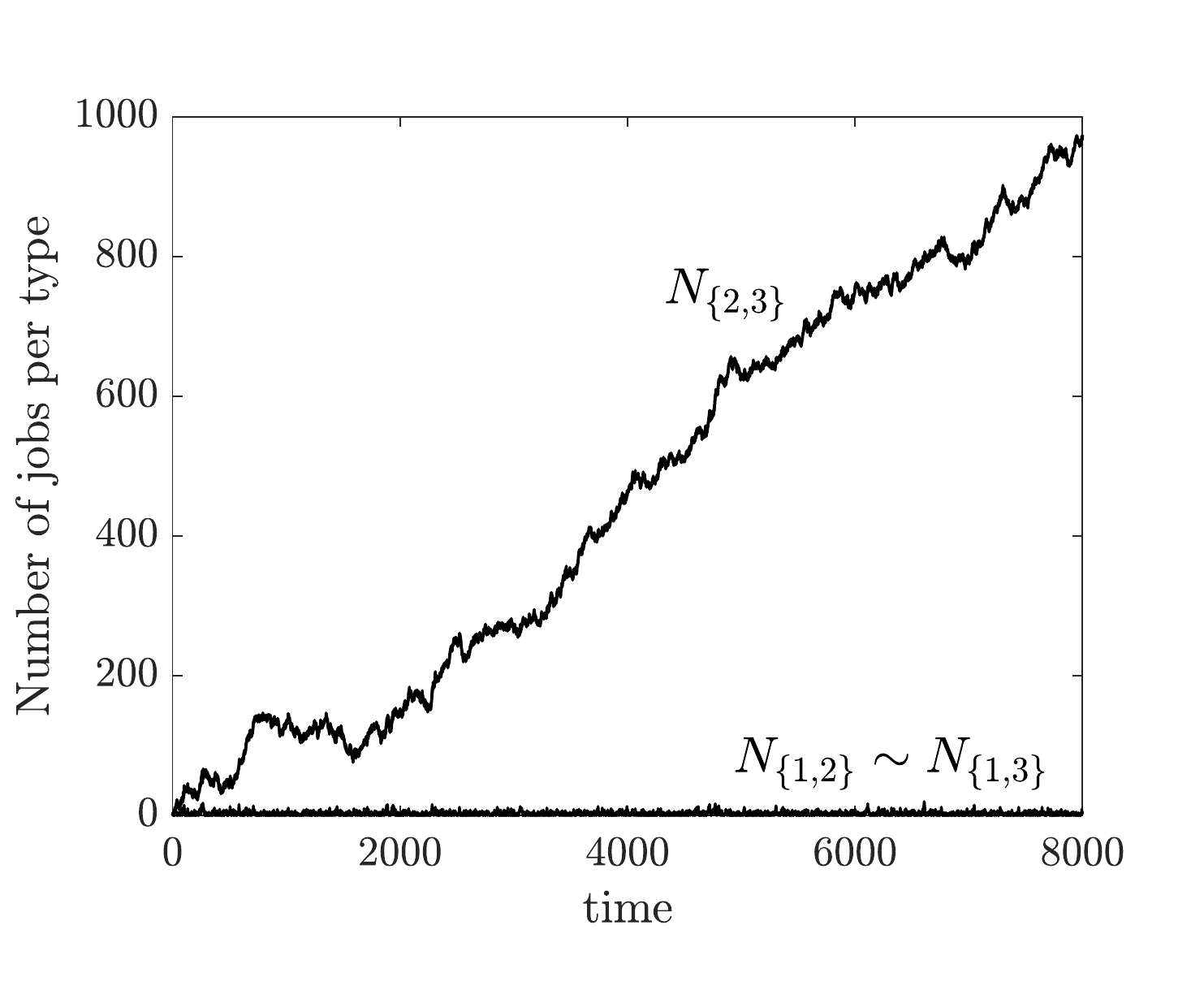}%priority_ok
		\caption{The trajectory of the number of jobs per type when  $\lambda=2.9$.}
		\label{ua:fig:pri}
	\end{center}		
\end{figure}

In Figure \ref{ua:fig:pri} we plot the sample-path of the number of jobs when 
%$\rho=0.96 <1$. 
$\lambda = 2.9 < 3 = \mu K$. 
We observe that   the number of type-$\{2,3\}$  jobs in the system grows large, while  the number of type-$\{1,2\}$ and type-$\{1,3\}$ jobs stay close to 0. Hence, the system is clearly unstable, even though $\lambda<\mu K$. This can intuitively be explained by the inefficiency  induced by the priority mechanism as  the type-$\{2,3\}$ jobs are preempted by type-$\{1,2\}$ and type-$\{1,3\}$ jobs in servers 2 and 3, respectively. We refer to~\cite{Anton2019} for more details.

%An intuitive explanation is a waste of resource due to the priority mechanism.
%This is explained as follows: type-$\{1,2\}$ jobs and  type-$\{1,3\}$ jobs are oblivious to the presence of  of type-$\{2,3\}$ jobs, due to the preemptive priority assumed in servers 2 and 3. Type-$\{1,2\}$ and type-$\{1,3\}$ jobs have  one server in common. Such a FCFS-redundancy system ($M$-model) has been analyzed in~\cite{Gardner16}, where it was obtained that this system (and hence the number of type-$\{1,2\}$ jobs and type-$\{1,3\}$ jobs) is stable when $\frac{\lambda}{3\mu}< \frac{3}{2}$. Which is satisfied in the present system. 

%Type-$\{2,3\}$ jobs are served in server 2 (3) whenever there are no type-$\{1,2\}$ jobs (type-$\{1,3\}$ jobs) present in the system. 
%Note that type-$\{1,2\}$ and type-$\{1,3\}$ jobs behave independent from type-$\{2,3\}$ jobs. Assuming type~$\{1,2\}$ and type~$\{1,3\}$ are in steady state, the  drift of the number of type-$\{2,3\}$ jobs in the system is given by 
%\begin{eqnarray}
%&&\frac{\mathrm{d}}{\mathrm{d}t}{\mathbb E}_{\stackrel{\rightarrow} n}\big[N_{\{2,3\}}(t)\big]\Big|_{t=0} = \frac{\lambda}{3} - 2\mu \left(\frac{(2\mu-\lambda/3)^2(3\mu - 2\lambda/3)}{4\mu^2(3\mu-2\lambda/3)+(\lambda/3)^2\mu}\right)\left(\frac{2\mu}{2\mu-\lambda/3} \right).\nonumber
%\end{eqnarray}
%It can be checked that the latter is strictly negative if and only if $\lambda < 0.91\mu K=2.73$.

\subsection{General Service Times}	
\label{ua:IID_gen}
To the best of our knowledge, no   stability results exist for general service times with i.i.d. copies. 
In this section, we present the stability result obtained for scaled Bernoulli service times, defined as $$ \left\{\begin{array}{ll}
X\cdot M, & \textrm{ with probability } 1/M \\
0, & \textrm{ with probability } 1-1/M,
\end{array}\right.
$$
where $M>0$ and $X$ is a strictly positive random
variable with $E[X] = 1$. In this setting, Raaijmakers et al. \cite{Raaijmakers2018} characterize the stability condition for the redundancy-$d$ model where FCFS is implemented and the number of servers grows large. 

\begin{proposition}[\cite{Raaijmakers2018}]
	Consider the redundancy-$d$ model under FCFS with  scaled Bernoulli service times and i.i.d. copies.  Then, $\lambda<\frac{M^{d-1}}{E[\min(X_1,\ldots,X_d)]}$ is a sufficient stability condition for any $M$. In addition, for any $\epsilon$, it holds that $(1-\epsilon)\lambda<\frac{M^{d-1}}{E[\min(X_1,\ldots,X_d)]}$ is a necessary condition, for $M$  sufficiently large.
\end{proposition}

We observe that the  stability condition is independent of the number of servers, but strongly depends on the number of copies $d$. The latter is in contrast to the exponentially distributed service times, where the stability condition does dependent on the number of servers but is independent of $d$ (see Proposition~\ref{ua:ps_ros_iid}). Thus, we observe that when copies are i.i.d., the stability condition strongly depends on the service time distribution. In addition, we observe that as $M$ grows large (and hence the variance of the service times grows large), the stability region increases by a factor $M^{d-1}$, by taking advantage of a greater diversity in service times. 

\section{Correlated Copies}
\label{ua:Cor}

% \cite{Gardner17b} : S&X model where jobs have identical copies / Homogeneous servers cap. 1 / genral service times / redundancy d / FCFS / RIQ algorithm. 
Several  studies (e.g., \cite{Vulimiri2012}) have shown that the i.i.d. copies assumption can be unrealistic, since large jobs remain large when replicated. Hence, having additional copies could lead to high response times and even instability. Motivated by the latter,  stability results with correlated copies have been the focus of recent literature.

\subsection{Identical Copies}
\label{ua:ident_cop}
% OR paer : Homogeneous serv /red-d/ iid copies / ROS-PS/ exponentials. 
%\rv{ Our contributions are here because it follows an chronological order, the papers that are cited later, where written later, except for Tim's first. Youri, Mendelson and Tim step a bit in our work so I thought it had more sense. }

% \cite{Raaijmakers2019a}: Delta proving with identical copies / FCFS / redundancy d / General service times / server capacities follow some joint distribution.
%\rv{I would remove this paper, does not fit much.} Under the identical copies assumption, Raaijmakers et al. \cite{Raaijmakers2019a} consider the redundancy-$d$ model with generally distributed server capacities where FCFS is implemented. The authors consider a system where each job dispatches $d$ identical probes (copies) to the relative servers and waits until the first to finish, moment when the job is assigned to the corresponding server and the additional probes are removed from the system. The authors obtain analytical results for delay performance measures in specific scenarios. 
%Moreover, the authors compare via simulation the mean latency of the system with known redundancy models. 
In this section, we assume that jobs have identical copies, i.e., all   copies belonging to one job have the same size. This correlation makes that a job can only depart due to its copy that has received most service so far. Thus, the instantaneous departure rate of a job depends on its copy that has currently attained most service. 

\textbf{FCFS Policy.}
With FCFS, the eldest job in the system will be served at all of its compatible servers. A job later in the queue will be served at its  compatible servers that are not engaged by earlier jobs in the queue.

%Let us consider the following descriptor: $\stackrel{\rightarrow} e=(l_{\ell^*},a_{\ell^* },c_{\ell^*},\ldots,c_2,l_1,a_1,c_1)$, similar to that in Section~\ref{ua:IID}. Here, ${\ell^*}$ denotes the number of jobs that receive service in state $ \stackrel{\rightarrow} e$ and $c_i$ denotes the type of the $i$-th job in service that is in service at servers $c_i\backslash c_1,\ldots,c_{i-1}$, and has attained service $a_i=(a_{i,s})_{s\in c_i}$, and $l_i$ denotes the number of jobs that arrived after job $c_i$ and have as compatible servers servers in $\cup_{j=1}^i c_j$, for $i=1,\ldots,\ell^*$. 

The stability condition for the redundancy-$d$ system with FCFS and exponentially distributed service times is characterized in Anton et al. \cite{Anton2019}, through the average departure rate per type in the so-called \emph{saturated system}. The latter assumes an infinite backlog of jobs waiting for service.
The long-run time-average number of jobs in service in the saturated system is denoted by $\bar\ell$.
%Hence, the total capacity of the saturated system is $\bar{\mathcal D}:=\sum_{c\in\mathcal C}\bar{\mathcal D}_c$. 
A detailed description of the saturated system and the characterization of $\bar\ell$ can be found  in \cite{Anton2019}.
\begin{proposition}[\cite{Anton2019}]\label{ua:Stab:fcfs:1}
	For the redundancy-$d$ system under FCFS with exponentially distributed service times and identical copies, the system is stable if   $\lambda < \bar\ell \mu$ and unstable if  $\lambda > \bar\ell \mu$.
\end{proposition}
The value of $\bar\ell$, and hence the stability region, can be numerically obtained by solving the balance equations of the saturated system, see \cite{Anton2019} for more details. We note that the instantaneous  departure rate in the saturated system strongly depends on the types in service.  As a consequence, no expression has been derived so far for $\bar\ell$ for general $K$ and $d$ values.

\begin{figure*}[th]
	\begin{minipage}{.5\textwidth}
		\centering
		\noindent
		\scriptsize
		\setlength{\tabcolsep}{1pt}
		\renewcommand{\arraystretch}{1.75}
		\begin{tabular}[b]{|c|c|c|c|c|c|c|c|}
			\hline
			$\bar\ell/K$	& $K=2$ & $K=3$ & $K=4$ & $K=5$ & $K=6$ & $K=7$ & $K=8$ \\
			\hline
			$d=1$ & 1	  & 1	  & 1	  & 1     &   1	 & 1 & 1 \\
			\hline
			$d=2$ & 0.5   & 0.66  &	0.71 & 0.74 &  0.76 & 0.77 & 0.77 \\     
			\hline
			$d=3$ &       & 0.33     &	0.5	  & 0.54  &   0.57 & 0.58 & 0.60 \\
			\hline
			$d=4$ &       &       &	0.25	  &  0.4	  &  0.43 & 0.46 &	0.47  \\
			\hline
			$d=5$ &       &       &		  &  0.2	  &   0.33   & 0.36 & 0.38   \\
			\hline
			$d=6$ &       &       &		  &   	  &  0.16 & 0.28 & 0.31  \\
			\hline
			$d=7$ &       &       &		  &  	  &    &  0.14  &	 0.25  \\
			\hline
		\end{tabular}
	\end{minipage}
	\hfill
	\begin{minipage}{.45\textwidth}
		\centering
		\includegraphics[width=1\textwidth]{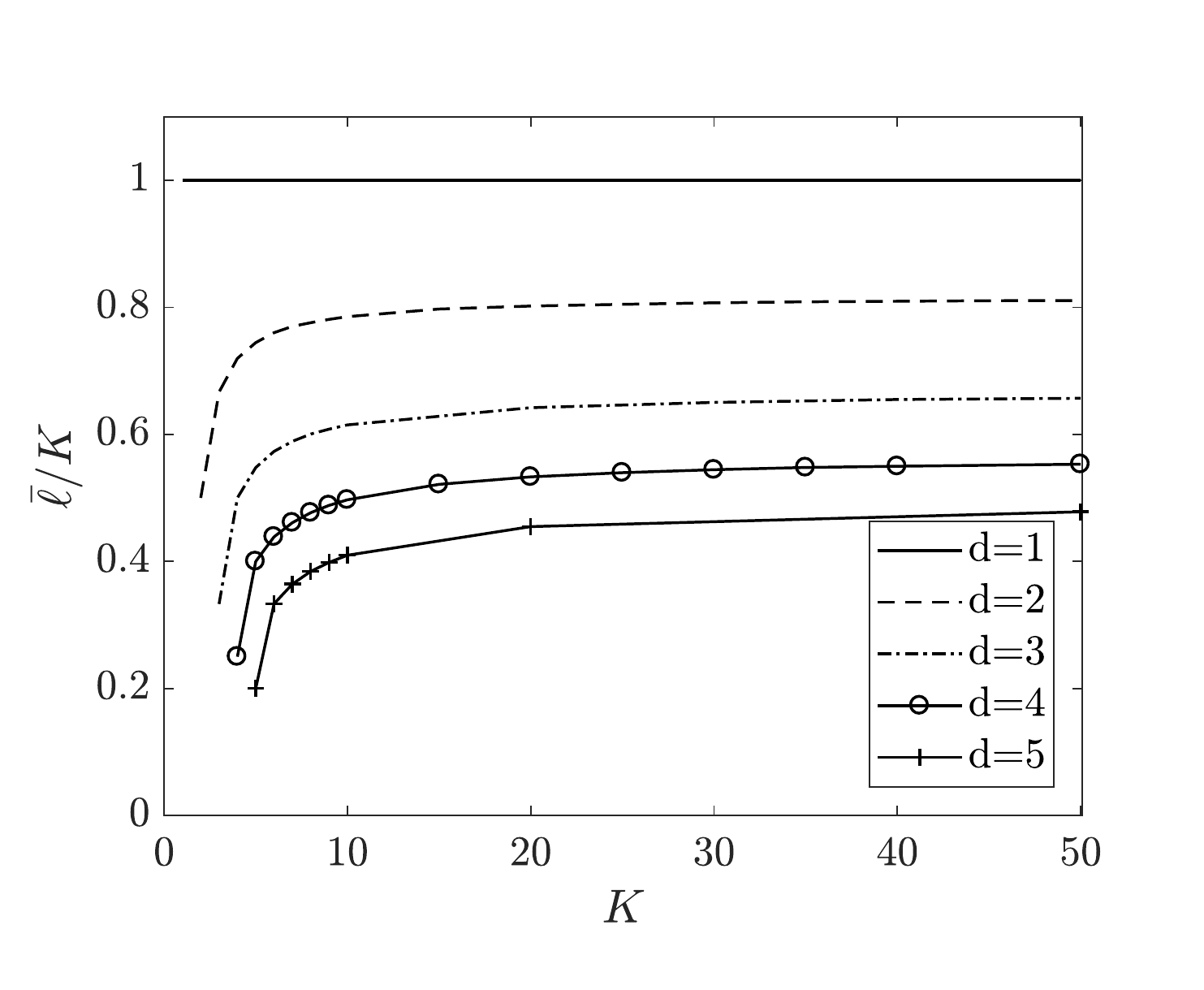}
	\end{minipage}
	\caption{The table and figure show the values of $\bar\ell/K$ for different values of $d$ and $K$.}
	\label{ua:tab:fig}
\end{figure*}

%%%%%%%%%%%%%%%%%%%%%%%%%%%%%%%%%%%%%%%%%%%%%%%%%%%%%%%%%%%%%%%%%%%%%
%Since jobs are dispatched uniformly among the servers,  it holds that $\bar{\ell}_c = \bar{\ell}/p_c$ where $\bar\ell$ is the total number of jobs in service in the saturated system. The stability condition thus reduces to $\lambda<\mu \bar{\ell}$, or equivalently,  $\frac{\lambda}{K\mu}<\frac{ \bar{\ell}}{K}$, where $\frac{\lambda}{K\mu}$ is the traffic load.

%Let $\bar E$ denote the state space of the saturated system. Then, the mean number of jobs in service can formally be written as
%\begin{equation}
%\label{C1:eq:averagesFCFS} 
%\bar\ell  := \sum_{\vec e\in\bar E} \pi(\vec e)\ell^*(\vec e),
%\end{equation}
%with
%$\ell^*(\vec e)=\ell^*((O_{\ell^*},\ldots,O_2,L_1,O_1))= \ell^*\mu$  the departure rate of the system when it is in state $\vec e\in\bar E$ and 
%$\pi(\vec e)$  the steady-state distribution of the saturated system.
%for $d\in \{1, K-1,K\}$.
Note that the  stability condition can equivalently be written as   $\frac{\lambda}{K\mu}<\frac{ \bar{\ell}}{K}$, where $\frac{\lambda}{K\mu}$ is the traffic load.
In Figure~\ref{ua:tab:fig} (originally in \cite{Anton2019}), we provide numerical values for $\frac{ \bar{\ell}}{K}$, that is, the traffic supported by the system.
The table (\emph{left}) shows $\bar\ell/K$ for small values of $K$ and the figure (\emph{right}) plots the value of $\bar\ell/K$ as $K$ grows large. To  obtain the value of $\bar\ell$ for $d\neq 1,K-2, K-1,K$, the authors simulate the saturated system, rather than solving the balance equations\footnote{When $d=K-1$, there are $d$ servers that process copies of one job, and the remaining $K-d=1$ server serves one additional job, hence, $\bar\ell=2$. 
	When instead $d=1$, there is no redundancy and each server serves one job in the saturated system, i.e., $\bar\ell=K$. When $d=K$, the system behaves as a single server with capacity $\mu$, that is, $\bar\ell=1$.}.
It was proven in \cite{Anton2019} that $\bar\ell/K$, hence the amount of supported traffic, increases when the number of servers ($K$) grows large, a property that can be observed in  Figure~\ref{ua:tab:fig}.

\textbf{Processor Sharing Policy.}
Under PS and identical copies, the stability condition is characterized in~\cite{Anton2020}. % for exponential distributed service times. 
There it is shown that the  stability condition coincides with that of a $K$ parallel server system where each type-$c$ job is only dispatched to its so-called least-loaded servers. In order to state this result, we first need to define several sets of servers and customer types.
The first subsystem includes all servers, that is $S_1=S$. 
We denote by $\mathcal L_1$ the set of least-loaded servers in the system~$S_1=S$. Thus, 
$$\mathcal L_1=\left\{s\in S_1\ : \ s=\arg\min_{\tilde s\in  S_1} \left\{\frac{1}{\mu_{\tilde s}}\sum_{c\in{\mathcal C(\tilde s)}} p_c\right\}\right\}.$$ 
For $i=2,\ldots, K$, we define recursively 
$$\begin{array}{rl}
S_i:= &S\backslash \cup_{l=1}^{i-1}\mathcal L_l,\\
{\mathcal C}_i:=& \{c\in\mathcal C \ : \ c\subset S_i\},\\
\mathcal C_i(s):=&\mathcal C_i\cap\mathcal C(s),\\
\mathcal L_i:=&\left\{s\in S_i \ : \ s=\arg\min_{\tilde s\in  S_{i}} \left\{\frac{1}{\mu_{\tilde s}}\sum_{c\in{\mathcal C}_{i}(\tilde s)} p_c\right\} \right\}.
\end{array}$$
The  $S_i$-subsystem refers to the system consisting of the servers in  $S_i$, with only jobs of types in the set ${\mathcal C}_i$. 
The $\mathcal C_i(s)$ is the subset of types that are served in server~$s$ in the $S_i$-subsystem. We let $ \mathcal C_1=\mathcal C$. The set  $\mathcal L_i$ represents the set of least-loaded servers in the $S_i$-subsystem. %, or in other words, the least-loaded servers in the $S_i$-subsystem.
Finally, we denote by $i^*:= \arg\max_{i=1,\ldots, K}\{\mathcal C_i \ : \ \mathcal C_i\neq\emptyset\}$ the last index $i$ for which the  subsystem $S_i$ is not empty of job types.

The stability condition is now characterized in \cite{Anton2020} by the least-loaded servers that can serve  each job type.% Denote by $\mathcal R(c)$ the set of  least-loaded servers that serve type-$c$ jobs: %, or in other words, the set of servers where  type-$c$ jobs achieve maximum capacity-to-fraction-of-arrival ratio: $$\mathcal  R(c):=\{s: \exists i, \mbox{ s.t. } c\in C_i(s) \mbox{ and } s\in \mathcal{L}_i\}.$$ Note that there is a unique  subsystem $S_i$ for which this happens, i.e., $\mathcal  R(c)\subseteq \mathcal{L}_i$ for exactly one $i$. We note that for a type-$c$ job, if $c$ contains at least a server that was removed in the $i$th iteration, then $\mathcal R(c)\subseteq \mathcal L_i$. We further let $\mathcal{R}:=\cup_{c\in \mathcal{C}}\mathcal R(c)$.

%\begin{corollary}\label{stab:cond2}
%	The redundancy system is stable if
%	$\lambda \sum_{c: s\in \mathcal{R}(c)} p_c <\mu_s$, for all $ s\in \mathcal R$.
%	%, with $\sum_{c: s\in \mathcal R(c)} p_c>0$.
%	The redundancy system is unstable if there exists an $s\in \mathcal R$ such that $\lambda \sum_{c: s\in \mathcal R(c)} p_c >\mu_s$.
%\end{corollary}

\begin{proposition}[\cite{Anton2020}]\label{ua:Stable:PS}
	Assume that the service time distribution is such that it has no atoms and is light-tailed in the following sense,
	\begin{equation}\label{ua:lt}\lim\limits_{r\to\infty} \sup_{a\geq 0} \mathbf E[(X-a)1_{\{X-a>r\}}\vert X>a]=0.\end{equation}
	For a redundancy system with a general topology under PS with identical copies, the system is stable if
	$\lambda \sum_{c\in C_i(s)}p_c <\mu_s$, for all  $s\in \mathcal{L}_i,$   $i=1,\ldots, i^*$.
	%, with $\sum_{c: s\in \mathcal R(c)} p_c>0$.
	The redundancy system is unstable if there exists $\iota \leq i^*$ and   $s\in \mathcal{L}_{\iota}$ such that $\lambda \sum_{c\in C_{\iota}(s)}p_c >\mu_s$. 
	%or equivalently,  $$\lambda \sum_{c: s\in c} p_c <\mu_s, \ \mbox{  for all } \ s=1,\ldots, K.$$
	%\textcolor{blue}{INSTEAD, COULD WE  WRITE .?}
\end{proposition}

%These technical conditions have been used previously in the literature to prove stochastic stability from fluid limits arguments (see \cite{Lee2008} and \cite{PFTA12}) in the context of processor sharing networks and cannot be avoided easily.
It can be seen (as observed in \cite{PFTA12}) that the light-tailed condition in \eqref{ua:lt} also implies
\begin{equation}\label{ua:lt:1}
\sup_{a\geq 0}  \mathbf E[(X-a)\vert X>a]\leq\Phi<\infty,
\end{equation} 
which is a usual light-tailed condition (see \cite{Foss2013}).
%also implies \eqref{eq:lt:2}. 
Hence, \eqref{ua:lt} and \eqref{ua:lt:1} exclude heavy tailed distributions like Pareto, but include large sets of distributions 
such as phase type (which are dense in the set of all distributions on $\mathbb R^+$), exponential and hyper-exponential distributions, as well as distributions with bounded support.

For the redundancy-$d$ model, the above stability condition simplifies into $\lambda<K \mu /d$. The latter coincides with the stability condition of a system where all the copies need to be served, that is,   the worst possible stability condition.

\textbf{ROS Policy.} When ROS is implemented in the servers,  it was shown in~\cite{Anton2019} that the stability condition is not reduced when adding redundant copies. This was proved for exponentially distributed service times and identical copies for the redundancy-$d$ model. However, as stated in Section~\ref{ua:Conclusions}, we believe that this holds true for any redundancy structure and any correlation structure.

\begin{proposition}[\cite{Anton2019}]\label{ua:ps_ros_ic}
	For the redundancy-$d$ model under ROS with exponentially distributed service times and identical copies, the system is stable if  $\lambda<K\mu$.
\end{proposition}

The intuition behind the above result is as follows. Whenever there are many jobs in a server, the probability that this server serves a copy of a job that has also a copy elsewhere in service will   be close to zero.  Hence, with a probability close to 1, all highly-loaded servers are serving copies of different jobs and their instantaneous departure rate equals  the sum of their capacities.

%Let us assume that there are multiple copies of the same job being served at various of its compatible servers. Due to the identical copies assumption, the departure rate of that job is characterized by the minimum among the service times of the copies in service. However, when the number of jobs in the system is large, the probability that more than one copy of the same job are simultaneously in service will be close to zero. Then, if a single copy of the job is in service, say in server-$s$, the instantaneous departure rate of the job is just $\mu_s$. Hence, if all busy servers are serving copies of different jobs (even if all the jobs are of the same type), the instantaneous departure rate is fully characterized as the sum of the busy servers  capacities.

%We note that the stability condition coincides with the one under i.i.d. copies. From the above intuition, and the stability condition under identical copies, 

%However, there is not known prove for that. 

\subsection{Generally Correlated Copies}
\label{ua:Stab:gen}
In this section, we consider redundancy models where the service times of the copies of each job are correlated according to some general structure.

%We first consider a system where servers have heterogeneous capacities distributed according to some finite discrete distribution and implement FCFS. We assume that jobs have general service times and identical copies, but experience different server speed variations (slowdowns) at each server, which are i.i.d. among serves and are sampled from a generic random variable. Hence, the service times of the copies at the different servers are correlated. We note that the present is a quite general workload model that subsumes the S\&X model (presented by \cite{Gardner17b}), when servers have constant speeds, and the identical copies model, when speed variations are constant and i.i.d. copies.

For FCFS, Raaijmakers et al. \cite{Raaijmakers2020} consider a general workload model, which   subsumes the S\&X model, introduced in \cite{Gardner17}. The main difference is that in \cite{Raaijmakers2020} the server capacities are not fixed, but each job samples  server capacities from a discrete and finite distribution. 
The authors assume that the  server speed variations (slowdowns)  are either distributed according to  New-Better-than-Used (NBU) or New-Worse-than-Used (NWU). See \cite{Ross1995} for more details on NBU and NWU distributions\footnote{
	%\begin{definition}\normalfont
	\footnotesize $X$ is said to be \emph{New-Better-than-Used (NBU)} if for all $t_1,t_2\in\mathbb R$, $ \bar F_X (t_1+t_2) \leq \bar F_X(t_1)\bar F_X(t_2).$ $X$ is said to be \emph{New-Worse-than-Used (NWU)} if for all $t_1,t_2\in\mathbb R$,  $\bar F_X (t_1+t_2) \geq \bar F_X(t_1)\bar F_X(t_2).$
	%\end{definition}
	%Let us denote by $R(x) = -\ln(\bar F_X(x))$ for $x\in\mathbb R$ the hazard function, and by $r(x)=R'(x)$ for $x\in\mathbb R$ the hazard rate function. That is, 
	%$$ r(x) = \frac{f(x)}{\bar F(x)} \textrm{ for } x\in\mathbb R. $$
	%We note that the hazard rate function describes the probability that a certain event happens in a fixed time interval in the future.
	A sufficient condition for $X$ to be NBU (NWU) is to have an increasing (a decreasing) hazard rate, i.e., $r(x)$ is increasing (decreasing) in~$x$.}.
%The authors consider the redundancy-$d$ model with heterogeneous arrival rates of types and obtain the following comparison result. 

Depending on the random variation in the server speed, the authors prove that either no replication ($d=1$) or full replication ($d=K$) provides a larger stability region. 
Note that here the stability region refers to a wider concept than what we considered before. That is, it refers to the set of arrival rates such that there exists a static assignment rule 
%\rv{(possibly depending on the various parameters of the model)}
that makes the system stable.

\begin{proposition}[\cite{Raaijmakers2020}]
	Consider the following model. Each job is routed to~$d$ servers according to some  static probabilistic assignment. Servers implement FCFS. Every time a server starts serving a new copy, it samples a speed variation, which is independent across servers.  The type of a job is determined by the capacities it would obtain in each server. A job has a generally distributed service time.
	%\textcolor{red}{ and identical copies?}. 
	\begin{itemize}
		\item 
		If the probabilistic assignment   can depend on the job type, and the speed variation follows an NBU distribution, then  the stability region for $d=1$   is larger or equal than  that for $d>1$.   
		
		\item If the probabilistic assignment  does not  depend on the job type, and the  speed variation follows an NWU  distribution, then the stability region for     $d=K$  is larger or equal than  that for  $d=1$.   
	\end{itemize}
\end{proposition}

%\textcolor{red}{every time a server starts serving a new copy, it samples a speed variation, which is independent across servers and follows either a NWU or NBY distribution, where the capacity of the server is determined by the type of the copy in service.}

%For NWU server speed variations, no such result exists. Instead, the authors in \cite{Raaijmakers2020} observe that when server capacities within a job type are balanced, no replication provides the largest stability region, and when server speeds are unbalanced, full replications provides the largest stability region. 
From the above we observe that the optimal redundancy degree does not depend on the job size distributions, but rather on the random variation in the server speeds for a given job among the servers.

A sufficient stability condition for the redundancy-$d$ model with FCFS has been obtained in Mendelson~\cite{Mendelson2020}.   He considers that 
%jobs have general service times with unit mean and 
the service times of the copies  $X_1,\dots,X_d$ are identically distributed with mean 1 and sampled from a joint  distribution $F(x_1,\ldots,x_d)$.
%\textcolor{red}{I removed this: (closed-form lower-bound on the stability condition. )}

\begin{proposition}[\cite{Mendelson2020}]
	Consider the redundancy-$d$ model where FCFS is implemented and the service times of the copies are sampled from a general joint distribution $F(x_1,\ldots, x_d)$.
	%Let $X_1,\ldots,X_d$ be the service times of the copies of one job. 
	Then, $\lambda<\lambda_{lb}$ is a sufficient stability condition, where 
	$$ \lambda_{lb} := \frac{\mu K}{\sum_{m=0}^d \left( \sum_{j=1}^{d-m} E[\min(X_1,\ldots,X_j)] +mE[\min(X_1,\ldots,X_d)] \right)P_m},$$
	and $P_m=\binom{K-d}{d-m}\binom{d}{m}/\binom{K}{d}.$
\end{proposition}

For the special cases $d=1$ and $d=K$,  the sufficient condition simplifies to $\lambda<\lambda_{lb}=K\mu$ and $\lambda<\lambda_{lb}=\mu/E[\min(X_1,\ldots,X_d)]$,
respectively, which are in fact also the necessary stability conditions.

%\rv{However, in general, this is not the case. 

%For example, when jobs have exponential service times and identical copies, the stability condition equals $\lambda<\bar\ell/ E[X]$, see Section~\ref{ua:ident_cop}.  However, the sufficient condition in~\cite{Mendelson2020} is given by $\lambda<\lambda_{lb}=K/(E[X]d)$, which is weaker since   $\lambda_{lb}\leq \bar\ell/ E[X]$.}

\ 

%That is, service speeds are sampled from a joint distribution for each job at each compatible server, which can be inter-dependent and non-identically distributed. Additionally, at each compatible server of the job, copies experience random speed variations which are i.i.d. sampled from some generic random variable. Thus,  This model also subsumes  the S\&X model presented by \cite{Gardner17b}.  

% \cite{Raaijmakers2020}: FCFS / heterogeneous servers follow some discrrte and finite distribution / red-d / speed variation servers / general NBU and NWU service distri / red-d / identical copies.

%

%\st{We assume that jobs have general service times and generally correlated copies, that is, the service times $X_1,\dots,X_k$ of the copies of one job can be sampled by a joint service time distribution $F(x_1,\ldots,x_k)$. }\textcolor{red}{Was this text also not applicable for MEndelson already?}

We now consider the redundancy-$d$ model where PS is implemented. Raaijmakers et al. \cite{Raaijmakers2019} characterize the stability condition under any service time distribution through the minimum of the service times of the copies of a job. The latter can be heuristically explained as follows: assume that all servers are equally loaded. Then, due to PS, the copy that completes first is the one with the smallest service time among all copies of the job.
%Since all servers are equally loaded under redundancy-$d$, Raaijmakers et al. \cite{Raaijmakers2019} can construct a lower-bound system where servers serve at constant speed and characterize the necessary stability condition  through the analysis of the fluid limit.

\begin{proposition}[\cite{Raaijmakers2019}]\label{ua:Stab_PS_g}
	For the redundancy-$d$ model under PS 
	where the service times of the copies are sampled from a general joint distribution $F(x_1,\ldots, x_d)$,   a necessary stability condition is \newline
	$\lambda d E[\min(X_1,\ldots,X_d)] <K\mu$. 
\end{proposition}

In the particular case where copies are identical, the authors in \cite{Raaijmakers2019} prove that Proposition~\ref{ua:Stab_PS_g} gives a sufficient and necessary stability condition, which is given by  $\lambda d <K\mu$.  We note that the latter coincides with the stability condition for light-tailed service times distributions provided in Proposition~\ref{ua:Stable:PS}.
%For the system where copies are i.i.d., and jobs have exponential service times, the stability condition reduces to $\lambda<K\mu$, which agrees with the condition in Proposition~\ref{ua:ps_ros_iid}. 
Moreover,  \cite{Raaijmakers2019} shows that the stability condition under NWU service time distributions, respectively NBU service time distributions, is larger, respectively smaller, than  that for exponential service times.

\section{Related Work}
\label{ua:realted}
In this section, we briefly overview relevant papers on redundancy. Even though the results do not deal directly with stability, they are important pointers for the reader who wishes to work on redundancy.

\subsection{Response Time}
The response time (a.k.a. delay) measures the time elapsed between arrival and departure. It is together with stability the main performance measure, and it has  received considerable attention. The first performance analysis of a redundancy  model  was for  \emph{cancel-on-complete} ($c.o.c.$) with exponentially distributed service times, independent and identically distributed (i.i.d.) copies and FCFS. As discussed in Section~\ref{ua:IID_exp}, Gardner et al. \cite{Gardner17,Gardner16} and Bonald and Comte \cite{Bonald17a} exploit the link between this redundancy system and the Order Independent queue~\cite{Krzesinski2011}, in order to show   that the steady-state distribution has a product form. The paper  \cite{Gardner17} showed that the mean response time  in the system reduces as the redundancy degree $d$ increases. 
Redundancy $c.o.s.$ with FCFS   and exponentially distributed job sizes has been analyzed in Ayesta et al. \cite{ABV18}, where it was shown that the steady-state distribution also has a product form. This was achieved by showing that this model fits within the framework of  multi-type jobs and multi-type servers studied in Visschers et al. \cite{Visschers2012}. 
The above results have motivated researchers to develop unifying frameworks to explain the emergence of product form distributions in redundancy models. This is done in  Ayesta et al. \cite{Ayesta2019} and  Gardner and Righter \cite{Gardner2020} by extending the frameworks of Visschers et al. \cite{Visschers2012} and Order Independent queues  \cite{Krzesinski2011}, respectively. 

Comte and Dorsman \cite{Comte2020} introduce the Pass-and-Swap queue, not included in the above unifying frameworks, but for which  the product-form of the steady-state distribution is preserved.  The authors provide several examples that fall into this framework, including  a loss variant of the $c.o.s.$ redundancy model.

The response time has also been studied in limiting regimes such as heavy traffic and mean field. Cardinaels et al. \cite{Cardinaels2020} consider both $c.o.c.$ and $c.o.s.$  
and establish that in heavy traffic  the joint distribution of the number of jobs of the various types converges to the product of an exponentially distributed random variable times a deterministic vector, a phenomenon known as state-space collapse.  
Hellemans et al. \cite{HvH18,Hellemans2019a} consider the mean-field regime and characterize the stationary workload distribution of $c.o.c.$ with FCFS, general service times and both identical and  i.i.d. copies.  In Hellemans et al. \cite{Hellemans2019} the authors generalize the previous result to other redundancy scheduling implementations such as replication if above certain threshold, delayed replication policy or replicate small jobs.  Another mean-field result can be found in Hellemans et al. \cite{Hellemans2018} where the authors analyze  the stationary response time and workload distributions of  JSW(d), JSQ(d) and redundancy-$d$ under  FCFS and general service times. 
%Lastly, in \cite{Hellemans2020} the authors study the mean response time in the mean field regime as the load approaches 1, i.e., in heavy traffic regime. 

%and observe that this is reduced \matt{compared to the i.i.d. case ?? or the absence of redundancy?}\rv{depends: homogeneous server red-d, both reduce, general only reduces compared to iid} due to adding redundant copies. 

\subsection{Optimizing Redundancy}
\label{ua:rel:opt}
The stability results   presented in this survey show that both the scheduling policy and the degree of redundancy can have a big impact on the stability region and hence on the performance of the system. 
%and we have also seen that the degree of redundancy, namely the number of copies created, also has an impact on the stability region.
Motivated by this, researchers have aimed at  \emph{i)} characterizing what is the optimal scheduling policy in the servers and \emph{ii)}  determining what is the optimum number of copies that should be created.

One of the first papers studying redundancy was by Koole and Righter \cite{KR08}, which  considered a system where jobs can dispatch i.i.d. copies  to any subset of servers in the system. The authors showed that with FCFS and NWU service time distributions, the best policy is to replicate to all the servers. 

Several optimality results have been derived for the Least-Redundant-First (LRF) scheduling policy, which  serves jobs in lower priority as their number of copies increases (jobs with the same number of copies are served according to FCFS). 
In particular, Gardner et al. \cite{Gardner2017,GHR19} consider  nested redundancy models with exponential service times and i.i.d. copies, and  show that the mean response time is minimized under LRF. We note that  a redundancy model is nested if for all $c,c'\in\mathcal C$, either \emph{i)} $c\subset c'$ or \emph{ii)} $c'\subset c$ or \emph{iii)} $c\cap c'=\emptyset$. 
%as the number of copies increases. 
%However, the maximum gain of adding redundant copies, is obtained with a little proportion of redundancy.
% Gardner et al. \cite{Gardner2017} consider the $W$-model and prove that LRF minimizes the mean response time. 

Akgun et al. \cite{Akgun2013} consider a two-server system in which  each server has  dedicated traffic, that is, each server is a unique compatible server for one job type. The authors consider %JSQ and JSW routing policies and 
the DCF (Dedicated-Customers-First) scheduling policy and analyze the efficiency and fairness for both dedicated and redundant jobs. 
%The authors provide a set of conditions under which each of the policies is efficient and fair, by mean of stochastically maximize the departure process and minimize the mean response time.

%Nageswaran et al. \cite{Nageswaran} consider the $N$-model   analyze the mean response time per type and characterize under which conditions redundancy under $c.o.c.$ and $c.o.s.$ is fair compared to the JSQ system, a system where there is no redundancy.

%In Koole and Righter \cite{KR08} the authors show that with FCFS and certain service time distributions (including exponential), the best policy is to replicate as much as possible.
%Raaijmakers et al. \cite{Raaijmakers2018}, consider FCFS  and consider non-exponential distributed service requirements and show that the stability region increases 
%in both the number of copies $d$, and in the parameter that describes the variability of the service requirements.
%\urtzi{can we say which distribution raaijmakers considers?}
%\matt{I commented above since we already mentionned this result}

Sun et al. \cite{Sun2016} consider various low-complexity redundancy scheduling techniques for systems where jobs have i.i.d. copies, and investigate when these are delay-optimal (or nearly-delay optimal) with respect to the stochastic ordering. These new scheduling techniques are based on job replication and job cancellation decision features. For instance, the authors show that the \emph{fewest unassigned task first with low-priority replication} and \emph{earliest due date first with replication} policies are nearly delay-optimal with NBU and NWU distributions, respectively.

\subsection{Related Models}
Redundancy as considered in this chapter is closely related to the $(n,k)$ fork-join system. In the latter,  %a job is encoded into $n$ blocks in such a way that $k$ out of $n$ blocks are sufficient to reconstruct the initial file. Such models have been studied by the coding theory community. In the $(n,k)$ fork-join system, 
there exist $n$ servers each one receiving one of the blocks, and the job is completed once $k<n$ blocks are served. If $k=1$, this model becomes equivalent to the redundancy-$n$ model with $c.o.c.$.

% Note that the $(n,k)$ fork-join system can be seen as the redundancy model where the first $k$ out of $n$ copies are requested, further when $k=1$, the model reduces to the redundancy model with i.i.d. copies. It is known as the $(n,k)$ fork-join system. 

For the $(n,k)$ fork-join model, 
%Joshi et al. \cite{Joshi15} derive the mean latency and the cost of computing resources for exponential service times and $k=1$ under $c.o.c.$ and $c.o.s.$, that is the redundancy-$K$ model. 
Lee et al. \cite{Lee17a} provide sequences of systems that upper and lower bound the original one, and that converge to the original system. Through these bounds, the authors characterize the mean response time of the system. Li et al. \cite{Li2016} derive that in the mean-field regime,  coding always improves the mean response time compared to the redundancy model, i.e., $(n,1)$.

%In \cite{Joshi15,Joshi2015,Joshi17}, the authors consider an $n$ server system  where FCFS is implemented under $(n,k)$ fork-join system and $(n,r,k)$ partial fork-join system, where the job is sent to $r$ out of $n$ servers uniformly chosen at random and waits to the first $k\leq r$ to complete.  In \cite{Joshi15,Joshi17} the authors consider a $(n,r,k)$ partial fork-join system and analyze effective replication strategies for various scenarios. %The authors show that for log-concave (low variable) service time distributions $c.o.s.$ redundancy reduces both latency and computing cost. %For the $(n,r,k)$ partial fork-join system the authors show that both latency and cost are minimized when $r$ increases for log-convex (high variable) service time distributions.

In \cite{Joshi17}, the authors consider the $(n,r,k)$ partial fork-join system, where the job is sent to $r$ out of $n$ servers uniformly chosen at random and waits for the first $k\leq r$ to complete. They study effective replication strategies for various scenarios. %The authors show that for log-concave (low variable) service time distributions $c.o.s.$ redundancy reduces both latency and computing cost. %For the $(n,r,k)$ partial fork-join system the authors show that both latency and cost are minimized when $r$ increases for log-convex (high variable) service time distributions.
The authors show that both latency and cost are minimized when $r$ increases for log-convex (high variable) service time distributions.  Duffy et al. \cite{Duffy2019}  compare  the tail response time of the $(n,r,k)$ model to that of the redundancy-$d$ model  (with   batch arrivals of size~$r$).
%and a single copy of each job is requested 
The authors show that the tail distribution of the response time under $(n,r,k)$ partial fork-join is smaller than under the redundancy-$d$ model as long as $r-k\geq d$, as the number of servers tend to infinity.

%\cite{Zubeldia2020} Homogeneous server system with S&X model (exponential slowdons), identical copies, k_n replicas out of all are needed to complete service . 

In a recent paper, Zubeldia \cite{Zubeldia2020} considers the $S\&X$ model where the slowdown experienced by each copy in service is independent across servers, but not necessarily independent from the job's service time. The author provides a lower-bound on the mean delay for the $(n,r,k)$ partial fork-join system, and shows that when slowdowns are exponentially distributed and independent of the service time of the job, the expected delay is minimized in the mean-field limit for a constant $r$ that only depends on the arrival rate and mean slowdown. %Furthermore, when the slowdown depends in some particular way in the service time of the job, the author shows that in the mean-field, the mean delay is minimized when smaller tasks are replicated more than larger tasks.

%mising papers  \cite{Wang2014,Poloczak16}.

\section{Conclusions, Open Problems and Conjectures}
\label{ua:Conclusions}
The literature on the stability analysis of redundancy systems is recent and growing. However, there are many important cases that have not been analyzed yet. In this  section, we address some of the open problems related to stability, and  state several conjectures that are based on our intuitive understanding of the system. It is our hope that this survey might encourage more research on this relevant and timely topic. 

%\subsection{I.i.d. copies}

\subsection{I.i.d. Copies.}
\label{ua:Stab_iid}
As shown in Proposition~\ref{ua:fcfs_iid}, FCFS is maximum stable with exponential service times and i.i.d. copies. We believe that this result should remain valid 
for any work-conserving scheduling policy with non-preferential treatment across types, for instance PS, ROS, LCFS, LAS and SRPT. 
The reason for this is that the i.i.d assumption combined with the non-preferential treatment across types permits to take advantage of diversity when the system is close to saturation.\\

%The stability region of the system where heterogeneous server redundancy model where jobs have exponentially distributed service times and i.i.d. copies. Under these assumptions and when the scheduling policy is work-conserving, the instantaneous departure rate for a given set of types present in the system is the maximum possible, that is, the sum of the capacities of all busy servers. 

%\begin{conjecture}\label{ua:IID:th}

\noindent\textbf{Conjecture 1.} Consider a redundancy system with a  general topology with exponentially distributed service times and i.i.d. copies. For any work-conserving non-preferential scheduling policy, the system is stable if for all $C\subseteq \mathcal C$, 
$$\lambda\sum_{c\in C}p_c < \sum_{s\in S(C)}\mu_s,$$
where $S(C)= \bigcup_{c\in C}  \{s\in c\}$.
%\end{conjecture} 
%\textcolor{red}{The below intuition is only for PS and ROs. Can we give instead an intuition that holds true for general policies??} The intuition behind the stability condition for either PS or ROS is as follows: The instantaneous departure rate of type-$c$ jobs, is given by $\sum_{s\in c}\mu_s\frac{N_c(t)}{M_s(t)}$, where $\mu_sN_c(t)/M_s(t)$ is the average capacity of server $s$ dedicated to type-$c$ jobs. Hence, for a given subset of types $C\subseteq\mathcal C$ present in the system, the arrival rate is $\lambda\sum_{c\in C}p_c$ and the instantaneous departure rate is $$ \sum_{c\in C} \sum_{s\in c}\mu_s\frac{N_c(t)}{M_s(t)} = \sum_{c\in C} \sum_{s\in c}\mu_s\frac{N_c(t)}{\sum_{c\in C : s\in c} N_c(t)} = \sum_{s\in S(C)} \mu_s.$$Hence, if $\lambda\sum_{c\in C}p_c<\sum_{s\in S(C)} \mu_s$, the total arrival rate to that subset of types is smaller than its departure rate, which allows to show that the system with type $C$ jobs is stable. 

\vspace{2mm}

\noindent
\textbf{Open Problem 1.}
If we relax the  exponential service times to general service time distribution, the
stability condition is unknown.

\subsection{FCFS Scheduling Policy with Identical Copies}

In Section~\ref{ua:ident_cop}, we saw that $\lambda/\mu K< \bar \ell/K$ is the stability condition of the redundancy-$d$ system where jobs have identical copies and exponential service times. 

\

\noindent 
\textbf{Open Problem 2.}
If we relax the redundancy-$d$ structure to general topologies, or the exponential service times to general service times, the stability condition is unknown. 

\ 

For exponential service times with the redundancy-$d$ structure,  we observed in  Figure~\ref{ua:tab:fig}  that for a given number of copies~$d$, $\lim\limits_{K\to\infty} \bar \ell/K <1$. Note that $\lambda /\mu K<1$ is the stability condition for a system with no redundancy. Hence, if it can be proved that  $\lim\limits_{K\to\infty} \bar \ell/K <1$, this would imply that   as the number of servers grows large, the traffic load that a redundancy system can support is smaller than if no redundancy was implemented. 

\

\noindent\textbf{Conjecture 2.} Consider the redundancy-$d$ model where FCFS is implemented and jobs have exponentially distributed service times and identical copies. Then, for fixed $d$, $\lim\limits_{K\to\infty} \bar \ell/K <1$.
\\ 

The limit should coincide with the stability condition given in \cite{HvH18}, where the authors develop a numerical method to derive the stability condition in the mean-field limit. 

We also observed the following monotonicity property in the number of redundant copies. More precisely, we conjecture that as the degree of redundancy increases, the stability region becomes smaller.\\

\noindent\textbf{Conjecture 3.} Consider the redundancy-$d$ model where FCFS is implemented and jobs have exponentially distributed service times and identical copies. Then, for fixed $K$, $\bar \ell$ is decreasing in $d$, and hence, the stability region is decreasing in~$d$.
%\end{conjecture}

\

%We did not succeed to obtain a prove for that conjecture. On the one hand, we searched for state-descriptors that fit in the framework of Order Independent queues, Quasi-reversible queues and Whittle networks, but could not succeed. This was motivated by the fact that under FCFS, the latter provided the steady-steady distribution and stability conditions of the system and by the similarities between the FCFS system and the PS and ROS systems.

%On the other hand, when describing the fluid-limit of the present system, we observe that for a given subset of types, the drift of the subsystem where only those types are present is negative. However, the drift of a state that transitions between two different subsystems, is discontinuous and does not necessarily rely inside a convex hull with negative drift. We note that this states are infinitely many.  

\subsection{ROS Scheduling Policy with Generic Correlation Structure.} 
In the particular case of ROS, we believe that Conjecture~1 will remain valid even if copies follow a general correlation structure, including identical copies. 
So far, this was only proved for the redundancy-$d$ model with  exponential distributed service times with identical copies, see   Proposition~\ref{ua:ps_ros_ic}.\\ 

%\begin{conjecture}\label{ua:ROSgen}
\noindent\textbf{Conjecture 4.}
Consider a redundancy system with a general topology with 
%Assume the heterogeneous redundancy system with $\mathcal C$ job types, 
exponentially distributed  service times and an arbitrary correlation structure among copies. ROS  is stable if for all $C\subseteq \mathcal C$, 
$$\lambda\sum_{c\in C}p_c < \sum_{s\in S(C)}\mu_s,$$
where $S(C)= \bigcup_{c\in C}  \{s\in c\}$.	 \\
%\end{conjecture}
\

The intuition would be the following. 
In principle, multiple copies of the same job could be served simultaneously at various of its compatible servers. Due to the heterogeneous capacities and the correlation among the copies, the departure rate of that job depends on the residual service time of each copy. 
%We note that if server capacities are homogeneous and copies are identical, the potential departure of a job is characterized by the minimum among the attained service times of the copies in service. 
However, when the number of jobs in the system grows large, the probability that more than one copy of the same job is simultaneously in service goes to zero. Using fluid-limit techniques,   as done in \cite{Anton2019}, one then obtains that the fluid limit of the system equals that of the system where jobs have i.i.d. copies. Hence, if Conjecture~1 is valid, this would imply that Conjecture~4 is true as well. \vspace{-3mm}

%%%%%%%%%%%%%%%%%%%%%%%%%%%%%%%%
\subsection{Redundancy-Aware Scheduling}
Another interesting, and so far unexplored area, is the impact of redundancy-aware scheduling policies on the stability region and  the performance of the system. By redundancy-aware we refer to policies like LRF or Most-Redundant-First that can use information on the number of copies when choosing which copy to serve in a server. As discussed in Section~\ref{ua:rel:opt}, the authors of \cite{Gardner2017,GHR19} consider the nested model with exponentially distributed service times and i.i.d. copies and show that LRF minimizes the mean response time. 
%This in particular implies that LRF is \emph{maximum stable}. 
It would be interesting to explore this further  for more general redundancy settings. \vspace{-3mm}

\section{Acknowledgement}
Research of E. Anton supported and research of M. Jonckheere partially supported by the French ``Agence Nationale de la Recherche (ANR)'' through the project ANR-15-CE25-0004 (ANR JCJC RACON). U.~Ayesta has received funding from the Department of Education of the Basque Government through the Consolidated Research Group MATHMODE (IT1294-19).
\vspace{-2mm}
%%%%%%%%%%%%%%%%%%%%%%%%%%%%%

%\bibliographystyle{plain}
%\bibliography{bibli}

\end{document}